\documentstyle[preprint,aps,psfig]{revtex}
\newcommand{\beq}{\begin{equation}}
\newcommand{\eeq}{\end{equation}}
\newcommand{\beqa}{\begin{eqnarray}}
\newcommand{\eeqa}{\end{eqnarray}}
\newcommand{\ba}{\begin{array}}
\newcommand{\ea}{\end{array}}

\begin{document}
\draft


\widetext
\title{Particles and Anti-Particles 
in a Relativistic Bose Condensate} 
\author{Luca Salasnich} 
\address{
Istituto Nazionale per la Fisica della Materia, 
Unit\`a di Milano Universit\`a,\\ 
Dipartimento di Fisica, Universit\`a di Milano, \\
Via Celoria 16, 20133 Milano, Italy\\
E-mail: salasnich@mi.infm.it}

\maketitle

\begin{abstract}
We study the Bose-Einstein condensation (BEC) 
for a relativistic ideal gas of bosons. 
In the framework of canonical thermal 
field theory, we analyze the role of particles and anti-particles 
in the determination of BEC transition temperature. 
At the BEC transition point we obtain two universal curves, 
i.e. valid for any mass value: 
the scaled critical temperature as a function of the scaled 
charge density of the Bose system, and the density ratio of 
anti-particles versus the scaled critical temperature. Moreover,  
we numerically investigate charge densities 
and condensed fraction 
ranging from the non-relativistic to the ultra-relativistic 
temperature, where analytical results are obtained. 
\end{abstract}


 \newpage 

\narrowtext 

\section{Introduction} 

Nowadays more than twenty experimental groups 
have achieved Bose-Einstein condensation (BEC) 
in clouds of confined alkali-metal atoms.$^1$ 
These studies have renewed the theoretical interest 
on non-relativistic$^2$ but also relativistic BEC. 
In the past years, relativistic BEC has been 
analytically investigated 
by several authors using Euclidean-time functional 
integration$^{3-7}$ but a quantitative numerical 
analysis with the temperature ranging form 
the non-relativistic to the ultra-relativistic 
limit has never been performed. 
\par
In this paper we consider BEC in the case of a 
relativistic non-interacting Bose gas 
described by a complex scalar field $\phi(x)$. 
We derive the exact equation of motion of $\phi(x)$ 
in the grand canonical ensemble of equilibrium statistical 
mechanics$^8$ without invoking functional integration$^9$ 
but using instead canonical field theory. 
By means of the Bogoliubov prescription$^{10}$, we write down 
the equations of the condensate order parameter and 
of thermal particles and anti-particles. 
We numerically investigate the effect of particles and anti-particles 
in the determination of BEC transition temperature $T_c$. 
Moreover we study the fraction of anti-particles 
in the system as a function of temperature 
and derive ultra-relativistic formulas for $T_c$ and 
the condensed fraction in a generic $d$-dimensional space. 

\section{Scalar Field and Legendre Anti-Transformation} 
 
The Lagrangian density of a non-interacting 
complex scalar field ${ \phi}(x)$ is given by 
\beq 
{ {\cal L}} =\left( \partial^{\nu}{ \phi}\right)^+ 
\left( \partial_{\nu} { \phi} \right) - m^2 { \phi}^+{ \phi} \; , 
\eeq 
where $m$ is the mass of the identical bosons described by the 
scalar field. 
To study the finite-temperature properties of a field-theory 
one needs the Hamiltonian ${ H}$ of the system. In our case, 
the canonical conjugate momentum $\Pi(x)$ of the scalar field 
$\phi(x)$ is 
\beq
\Pi = {\partial {\cal L} \over \partial {\dot \phi} } 
= {\dot \phi}^+  \; , 
\eeq
and the Hamiltonian density reads 
\beq 
{ {\cal H}} = {\Pi}^+ {\Pi} + 
\nabla { \phi}^+ \cdot \nabla { \phi} 
+ m^2 { \phi}^+ { \phi} \; . 
\eeq 
Note that the invariance of the system 
under a global gauge U(1) field transformation 
implies the conservation of the electric current and the 
conserved charge density is 
\beq
{\cal Q} = i \left( \phi^+ {\dot \phi} - 
\phi {\dot \phi}^+ \right) = 
i \left(\Pi^+ \phi^+ - \Pi \phi \right) \; . 
\eeq 
\par 
Equilibrium statistical mechanics tells us that 
the grand canonical partition function $Z$ of 
a quantum system of Hamiltonian $H$ and 
conserved charge $Q$ is given by 
\beq
Z = Tr\left[ e^{-\left({ H} - \mu { Q} \right)/T} 
\right] \; , 
\eeq
where $T$ is the temperature 
of the thermal reservoir and $\mu$ is the chemical potential.$^8$ 
Thus the system in the grand canonical ensemble 
is described by an effective Hamiltonian density 
${\cal H}_{\mu} = {\cal H} - \mu {\cal Q}$, namely 
\beq
{\cal H}_{\mu} = \Pi^+ \Pi + \nabla \phi^+ \cdot 
\nabla \phi + V(\phi) - i \mu (\Pi^+ \phi^+ - \Pi \phi ) \; . 
\eeq
To find the effective Lagrangian density we observe that 
the Hamilton equation for ${\dot \phi}$ is given by 
\beq
{\dot \phi} = {\partial {\cal H}_{\mu} \over \partial \Pi} = 
\Pi^+ + i \mu \phi \; .  
\eeq 
Using the Legendre anti-transformation one can easily 
obtain the effective Lagrangian density 
${\cal L}_{\mu}= {\dot \phi}^+\Pi^+ + {\dot \phi}\Pi 
- {\cal H}_{\mu}$, namely  
\beq 
{\cal L}_{\mu} = {\cal L} + i \mu 
\left( \phi^+ {\dot \phi} - 
\phi {\dot \phi}^+ \right) + \mu^2 \phi^+ \phi \; . 
\eeq
Therefore, the introduction of the chemical potential 
$\mu$ is equivalent to the use of an effective Lagrangian 
${\cal L}_{\mu}$, that can be obtained with a shift 
\beq 
{\partial \over \partial t} 
\to {\partial \over \partial t} + i \mu \; ,  
\eeq 
in the time partial derivative of the bare Lagrangian ${\cal L}$. 
It is important to observe that 
the same result could be obtained by means of the 
Euclidean-time functional integration$^9$ 
and that the shift found holds also in the fermionic case.$^{11}$ 
Finally, the Euler-Lagrange equation of the effective 
Lagrangian density ${\cal L}_{\mu}$ reads 
\beq
\left[\Box + 2 i \mu {\partial \over \partial t} - \mu^2 
+ m^2 \right] { \phi} = 0 \; ,
\eeq
where $\Box={\partial^2 \over \partial t^2} - \nabla^2$ is the 
d'Alambert operator. 

\section{Bose-Einstein Condensation and Bogoliubov Prescription} 

In a Bosonic system one can separate Bose-condensed 
particles from non-condensed ones by means of the 
Bogoliubov prescription$^{10}$ that is given by 
\beq
\phi = \Phi + \eta  \; ,
\eeq
where 
\beq
\Phi=\langle \phi\rangle = {1\over Z} 
Tr\left[ { \phi} \; e^{\beta \left({ H} 
- \mu { Q} \right)} \right] \; , 
\eeq 
is the order parameter of the Bose condensate 
(a classical complex scalar field), namely the non-vanishing 
thermal average of the Bosonic field, 
and $\eta(x)$ is the operator of the non-condensed or thermal 
particles, such that $\langle\eta\rangle=0$. 
\par
The exact equation of motion of the order parameter $\Phi(x)$ 
is obtained by calculating the thermal average over the 
equation of motion of the scalar field $\phi(x)$. 
If we have a static and homogeneous order parameter, then  
\beq 
\left[- \mu^2 + m^2 \right] \Phi = 0 \; , 
\eeq 
from which it follows that there is 
macroscopic occupation of the lowest single-particle 
state ($\Phi\neq 0$) only if $|\mu| =  m$. 
\par 
The exact equation of motion of the fluctuation operator 
$\eta(x)$ is easily obtained by subtracting the 
exact equation of $\Phi(x)$ to the equation 
of $\phi(x)$. Note that, in the non-interacting case, 
the equations of $\phi(x)$, $\Phi(x)$ and $\eta(x)$ 
are formally identical but they have different meanings. 
In addition, by using the Bogoliubov prescription and 
the effective Lagrangian, the grand canonical 
charge density of the system reads 
\beq 
{\cal Q} = {\cal Q}_0 + {\tilde {\cal Q}} \; , 
\eeq 
where 
\beq
{\cal Q}_0 = 2 \mu |\Phi|^2 
\eeq
is the contribution due to the Bose 
condensate and 
\beq 
{\tilde {\cal Q}} = i \left( \eta^+ {\dot \eta} -
\eta {\dot \eta}^+ \right) 
\eeq 
is the contribution due to thermal particles and anti-particles 
(see also Ref. 3). 

\section{Bose Condensate and Thermal Particles} 

The relativistic complex scalar field operator 
$\eta(x)$ satisfies the equal-time commutation rule 
\beq
[\eta ({\bf x},t),\eta^+({\bf y},t)]=
\delta^3({\bf x}-{\bf y}) \; . 
\eeq 
The operator $\eta(x)$ 
can be Fourier decomposed into a single-particle basis 
of particles and anti-particles 
\beq 
\eta(x) = \sum_{\bf k} 
\left( 
{e^{i ({\bf k}\cdot {\bf x}-\omega_k t)} 
\over \sqrt{2\omega_k V} } a_k 
+ {e^{-i ( {\bf k}\cdot {\bf x} - {\bar \omega}_k t) } 
\over \sqrt{2{\bar \omega}_k V} } b_k^+ \right) \; ,
\eeq
where we have used the symbols $\omega_k$ and ${\bar \omega}_k$ 
to indicate the value of energy for particles and anti-particles, 
respectively. The Bose operators for particles 
and anti-particles satisfy the canonical 
commutation relations 
\beq  
[a_{\bf k},a_{{\bf k}'}^+]=[b_{\bf k},b_{{\bf k}'}^+]
=\delta_{{\bf k}{\bf k}'} \; ,
\eeq
and all other commutators are zero. 
In addition, one imposes the following Bose-Einstein 
thermal averages 
\beq 
\langle a_{\bf k}^+ a_{{\bf k}'}\rangle = 
{1\over e^{\omega_k/T} - 1 } \delta_{{\bf k}{\bf k}'} \; , 
\;\;\;\; 
\langle b_{\bf k}^+ b_{{\bf k}'}\rangle = 
{1\over e^{{\bar \omega}_k/T} - 1 } \delta_{{\bf k}{\bf k}'} \; . 
\eeq
The energies $\omega_k$ and ${\bar \omega_k}$ 
are determined by inserting the decomposition of 
the field $\eta(x)$ in its equation of motion, namely Eq. (10) with 
$\eta(x)$ instead of $\phi(x)$. 
In this way, one finds two decoupled algebric equations:  
\beq
\omega_k^2 - 2\mu \omega_k + \mu^2 - m^2 - k^2  = 0 \; , 
\eeq
\beq
{\bar \omega}_k^2 + 2\mu {\bar \omega}_k + \mu^2 - m^2 - k^2 = 0 \; ,  
\eeq 
which give the physical solutions 
\beq
\omega_k = \sqrt{k^2 + m^2} - \mu \; , 
\;\;\;\; 
{\bar \omega}_k = \sqrt{k^2 + m^2} + \mu \; . 
\eeq 
The Fourier decomposition and the energies $\omega_k$ 
and ${\bar \omega}_k$ enable us to calculate the 
thermal average ${\tilde q}=\langle 
{\tilde {\cal Q}}\rangle$ of the non-condensed 
charged density, which is given by 
\beq 
{\tilde q} = 
\sum_{{\bf k}\neq {\bf 0}} 
\left[ {1\over e^{( \sqrt{k^2+m^2} - \mu)/T} -1} 
- {1\over e^{( \sqrt{k^2+m^2} + \mu )/T} -1} 
\right] \; . 
\eeq 
Thus, ${\tilde q}=n_1-n_2$, 
where $n_1=\sum_{\bf k} \langle a_{\bf k}^+ a_{\bf k} \rangle$ 
is the average density of particles 
and $n_2=\sum_{\bf k} \langle b^+_{\bf k} b_{\bf k} \rangle$ 
is the average density of anti-particles. 
Note that $n_2$ is obtained 
from $n_1$ with the substitution $\mu \to - \mu$. 
The chemical potential $\mu$ 
describes both bosons and antibosons: the sign of $\mu$ 
indicates whether particles outnumber antiparticles or vice versa. 
Moreover, because both $n_1$ and 
$n_2$ must be positive definite, 
it follows that $|\mu |\le m$. 
\par
Obviously, the total number of 
particles is not conserved because of the production 
of antiparticles, which becomes relevant when $T$ 
is comparable with $m$. The conserved quantity 
is the net charge density $q=q_0+{\tilde q}$, where 
$q_0={\tilde Q}_0=2\mu |\phi|^2$ is the condensed charge 
density and ${\tilde q}=\langle {\tilde Q}\rangle=
n_1-n_2$ is the the difference between the density 
of particles and the density of anti-particles. 
The condensed charge density $q_0$ is non-zero 
only below the BEC transition temperature $T_c$. 
The condensed charge density corresponds to $k=0$ in Eq. (24) 
and it is thus given by 
\beq
q_0 =  {1\over e^{( m - \mu)/T} -1} 
- {1\over e^{( m + \mu )/T} -1}  = n_1^{(0)} - n_2^{(0)} 
\; , 
\eeq
where $n_1^{(0)}$ and $n_2^{(0)}$ are the 
density of condensed particles and anti-particles, respectively. 
It is easy to show that 
\beq 
\lim_{T\to 0} n_1^{(0)}(T)= q_0 \; , \;\;\;\; 
\lim_{T\to 0} n_2^{(0)}(T)= 0 \; . 
\eeq
As expected, at $T=0$ the particles 
are all in the condensate and there are no anti-particles. 
Moreover, in the limit $T\to 0$ the asymptotic behavior 
of the chemical potential $\mu$ and of the 
density $n_2^{(0)}$ of condensed anti-particles read 
$$
\mu(T)\sim m - T \ln{\left({q_0+1\over q_0}\right)}  
\; , 
$$
\beq
n_2^{(0)}(T) \sim {q_0+1\over q_0 \left(e^{2m/T}-1\right) -1} \; . 
\eeq 
To our knowledge, this is the first paper where these simple 
asymptotic relations have been explicitely written down. 

\section{Numerical and Analytical Results} 

The behavior of the Bose gas ranging from the non-relativistic 
to the ultra-relativistic regime can be numerically 
investigated by means of the non-condensed thermal charge 
density ${\tilde q}$ given by Eq. (24). 
In particular, we work in the thermodynamic limit substituting 
the sums in Eq. (24) with integrals and find  
\beq 
{\tilde q} = {1\over 2 \pi^2} \int_0^{\infty}dk \left[ 
{1\over e^{( \sqrt{k^2+m^2} - \mu)/T} -1} 
- {1\over e^{( \sqrt{k^2+m^2} + \mu )/T} -1} \right] 
= {1\over 2 \pi^2} \int_0^{\infty} dk 
\left[ n_1(k) - n_2(k) \right] \; , 
\eeq 
where $n_1(k)$ and $n_2(k)$ are the density profiles in momentum 
space for particles and antiparticles, respectively. 
It is important to stress that the previous formula has 
an useful scaling property: the chemical potential, 
the temperature and the momentum 
can be measured in units of $m$ and the densities in units 
of $m^3$. This follows from the fact that, with $\hbar=c=1$,  
the mass, the chemical potential, 
the temperature and the momentum have the same unit: the energy, 
while the length is measured in units of the inverse of energy. 
\par 
We first consider the case $T>T_c$. 
Given the charge density and the temperature,  
the chemical potential is fixed by Eq. (28). In such a way 
one determines also the fraction of anti-particles in the 
system. As shown in Fig. 1, where we plot $n_2/n_1$ as a function of 
the scaled temperature $T/m$ for different values of the 
scaled charge density $q/m^3$ of the Bose system, 
one can identify two regimes: 
the non-relativistic regime ($T/m\ll 1$) 
and the ultra-relativistic regime ($T/m\gg 1$). 
In the non-relativistic regime, the fraction 
of anti-particles of the system is negligible. 
In the ultra-relativistic regime the fraction of anti-particles 
becomes relevant. The curves with fixed charge density 
$q={\tilde q}$ ($q_0=0$) ends at the scaled critical 
temperature $T_c/m$, where the Bose-Einstein condensate appears. 
The role of temperature in the formation of 
anti-particles is also shown in Fig. 2, where we plot 
the density profiles in momentum space of particles and anti-particles 
for three increasing values of the scaled temperature. 
\par 
We observe that in the ultra-relativistic regime 
one can derive analytical results by performing 
a Taylor expansion of ${\tilde q}$ at first order in $\mu$. 
After straightforward but tedious calculations one finds  
\beq 
n_1 = {\zeta(3)\over \pi^2}T^3 + {\mu \over 6}T^2 \; , 
\;\;
n_2 = {\zeta(3)\over \pi^2}T^3 - {\mu \over 6}T^2 \; , 
\;\;
{\tilde q}={\mu \over 3} T^2 \; ,  
\eeq 
where $\zeta(x)$ is the Riemann $\zeta$-function.  
These analytical results, confirmed by our numerical calculations, 
show that although $n_2/n_1\to 1$ as $T\to \infty$, 
the charge density ${\tilde q}=n_1-n_2$ goes to infinity. 
\par 
We have previously shown that 
the critical temperature $T_c$ at which BEC occurs corresponds 
to $|\mu| = m$. At the BEC transition temperature $T_c$, 
the thermal charged density ${\tilde q}$ 
can still be determined from Eq. (28). In fact,  
by inverting the function ${\tilde q}(T_c,m=\mu)$ one finds 
the transition temperature. 
In Fig. 3 we plot two curves which do not depend on the 
value of the mass of Bosons in the gas (we call them universal curve). 
The first unversal curve is the scaled critical 
temperature $T_c/m$ as a function of the scaled charge density 
$q/m^3$. The second universal curve is 
the ratio $n_2/n_1$ between anti-particles 
and particles as a function of the scaled critical 
temperature $T_c/m$. As expected, $T_c$ grows with $q$ 
and $n_2/n_1$. Moreover, at a fixed density ratio $n_2/n_1$, 
it is easier to get high-temperature BEC 
with heavy-mass particles. The two universal curves 
of Fig. 3 can be compared with ultra-relativistic 
analytical results. From (29) one immediately finds 
that the critical temperature ($\mu=m$) is given by  
\beq
T_c = \left({3 q\over m}\right)^{1/2} \; ,
\eeq 
and the density ratio reads 
\beq
{n_2\over n_1} = {{\zeta(3)\over \pi^2}T_c^3 - {\mu \over 6}T_c^2
\over 
{\zeta(3)\over \pi^2}T_c^3 + {\mu \over 6}T_c^2 } \; . 
\eeq 
Fig. 3 shows that while 
the formula of the density ratio $n_2/n_1$ 
is valid only in the ultra-relativistic region 
(for $T_c\to 0$ it predicts the wrong limit $n_2/n_2\to -1$), 
the formula of the critical temperature is quite accurate 
also at low temperatures. 
Note that the formula of the critical 
temperature has been first obtained by Kapusta.$^3$ 
We now extend it to the case of a ultra-relativistic gas 
in $d$-dimensional space. 
\par
The charge density our system of non-interacting 
bosons can be re-written as 
\beq 
{\tilde q} =\int_0^{\infty} 
d\epsilon \; \rho(\epsilon ) \; 
\left[ {1\over e^{( \epsilon - \mu)/T} -1} 
- {1\over e^{( \epsilon + \mu )/T} -1} \right] \; ,
\eeq 
where $\rho (\epsilon )$ is the density of states. 
It can be obtained from the formula 
\beq 
\rho(\epsilon ) = \int {d^dk \over (2\pi)^d} 
\delta (\epsilon - H(k)) \; ,  
\eeq
where $H(k)$ is the classical single-particle 
Hamiltonian of the system in a $d$-dimensional space. 
The classical single-particle Hamiltonian of a relativistic ideal gas is 
$H=\sqrt{k^2 + m^2}$ and the density of states reads 
\beq
\rho(\epsilon )={2\pi^{d/2} \over (2\pi)^d \Gamma(d/2)} 
\epsilon (\epsilon^2-m^2)^{(d-2)/2} \; . 
\eeq 
In the ultra-relativistic limit the density of states is simply 
$\rho(\epsilon )=(2\pi^{d/2})/((2\pi)^d \Gamma(d/2)) 
\epsilon^{(d-1)}$. In this case, by 
using again the Taylor expansion of $q$ at first order 
in $\mu$ with $T=T_c$ one finally obtains 
\beq 
T_c = \left( { (2\pi)^d \Gamma(d/2)  
\over 4\pi^{d/2} \Gamma(d) \zeta(d-1)} 
{q \over m} \right)^{1/(d-1)} \; , 
\eeq 
where $\Gamma(x)$ is the factorial function. 
Because $\zeta(1)=\infty$, it follows that 
for a homogeneous relativistic gas 
there is BEC only for $d>2$, as in the case of a 
non-relativistic homogeneous gas (see also Ref. 12, Ref. 13).  
\par 
Below $T_c$, a macroscopic number of particle occupies 
the single-particle ground-state of the system 
($q_0\neq 0$). The Eq. (28) 
gives the charge density ${\tilde q}=q-q_0$ of non-condensed 
particles. In this way, form Eq. (15), one determines 
the order parameter $\Phi$, that is such that 
\beq 
|\Phi|^2 = {{\tilde q}(T_c) - {\tilde q}(T) \over 2 m} \; . 
\eeq
Thus, the condensed fraction $q_0/q$ can be 
numerically calculated as 
$1-{\tilde q}(T)/{\tilde q}(T_c)$. 
In Fig. 4 we show the condensed fraction as a function 
of the temperature for different values of the scaled 
charge density $q/m^3$ of the Bosonic particles. 
In the ultra-relativistic regime, from (28) and (29) 
one finds  
\beq
{q_0\over q} = 1 - \left({T\over T_c}\right)^2 \; , 
\eeq
namely the condensed fraction has an inverted-parabola shape. 
It is important to stress that, although we are able to determine 
the charge density $q_0$ of the Bose condensate, 
our formalism cannot tell us the fraction 
of anti-particles into the condensate. 
\par
For an ideal gas of charged massless Bosons 
($m=0$) it follows, form Eq. (30) and Eq. (35), that 
$T_c=\infty$ and $q_0=q$: at any temperature, 
all net charge resides in the Bose condensate. 
Nevertheless, if the thermal average of the 
charge is not conserved ($\mu =0$), 
i.e. a gas of photons, then BEC does not take place. 
Finally, by using the previously discussed procedure, 
one finds that the Bose condensed fraction 
for a ultra-relativistic gas in a d-dimensional space 
is given by  
\beq
{q_0\over q} = 1 - \left( {T\over T_c} \right)^{d-1} \; ,  
\eeq 
remembering that $T_c\to \infty$ as $d\to 2$. 

\section{Conclusions}

We have studied thermal properties 
of a non-interacting relativistic Bose gas 
by analyzing in detail the fraction of anti-particles 
in the system. By using a finite-temperature operator formalism, 
we have obtained the equation of the Bose condensate, 
described by a complex classical order parameter, 
and the equation of non-condensed particles and 
anti-particles. At zero temperature the particles 
are all in the condensate and there are no anti-particles. 
In the limit of zero temperature we have determined 
the asymptotic behavior of the density of condensed 
anti-particles. The charge density and 
the density of particles and anti-particles 
have been analyzed as a function of temperature 
raging from the non-relativistic to 
the ultra-relativistic regime. We have determined two universal 
curves at the BEC transition point: 
the scaled critical temperature $T_c/m$ as a function 
of the scaled charge density $q/m^3$ of the Bose gas, 
and the ratio between anti-particles and particles 
as a function of the scaled critical temperature $T_c/m$. 
Moreover, we have investigated the condensed fraction as 
a function of the scaled temperature for increasing values 
of the scaled charge density of the gas. 
Finally, analytical results have been found in the 
ultra-relativistic region. Our analytical formulas 
for a ultra-relativistic Bose gas in d-dimensional space 
generalize previous findings with $d=3$.  
\par 
In conclusion, we observe that detailed analytical and numerical 
investigations can be also performed in 
the case of an interacting relativistic Bose gas, 
at least in the Bogoliubov-Popov mean-field approximation. 
This is one of our future projects. 

\newpage 

\section*{References}

\begin{description}

\item{\ 1.} M.H. Anderson, {\it et al.}, 
Science {\bf 269}, 189 (1995); 
K.B. Davis, {\it et al.} Phys. Rev. Lett. {\bf 75}, 
3969 (1995); C.C. Bradley, {\it et al.},  
Phys. Rev. Lett. {\bf 75}, 1687 (1995). 

\item{\ 2.} F. Dalfovo, S. Giorgini, L.P. Pitaevskii, 
and S. Stringari, Rev. Mod. Phys. {\bf 71}, 463 (1999). 

\item{\ 3.} J.I. Kapusta, Phys. Rev. D {\bf 24}, 426 (1981). 

\item{\ 4.} H.E. Haber and H.A. Weldon, Phys. Rev. Lett. {\bf 23}, 
1497 (1981); H.E. Haber and W.A. Weldon, 
Phys. Rev. D {\bf 25}, 502 (1982) 

\item{\ 5.} J. Bernstein and S. Dodelson  
Phys. Rev. Lett. {\bf 66}, 683 (1991). 

\item{\ 6.} D.J. Toms, Phys. Rev. Lett. {\bf 69}, 1152 (1992); 
D.J. Toms, Phys. Rev. D {\bf 50}, 6457 (1994). 

\item{\ 7.} K. Shiokowa and B.L. Hu, Phys. Rev. D {\bf 60}, 105016 (1999). 

\item{\ 8.} K. Huang, {\it Statistical Mechanics} 
(John Wiley, New York, 1987).

\item{\ 9.} J.I. Kapusta, {\it Finite Temperature Field Theory} 
(Cambridge Univ. Press, Cambridge, 1989); 
M. Le Bellac, {\it Thermal Field Theory} 
(Cambridge Univ. Press, Cambridge, 1996). 

\item{\ 10.} N.N. Bogoliubov, J. Phys. U.S.S.R. {\bf 11}, 23 (1941); 
S.T. Beliaev, Sov. Phys. JEPT {\bf 7}, 289 (1958). 

\item{\ 11.} M. Modugno, Rivista del Nuovo Cimento {\bf 23}, N.5, 1 (2000). 

\item{\ 12.} L. Salasnich, Int. J. Mod. Phys. B {\bf 14}, 405 (2000). 

\item{\ 13.} L. Salasnich, J. Math. Phys. {\bf 41}, 8016 (2000).  

\end{description}

\newpage 

\begin{figure}
\centerline{\psfig{file=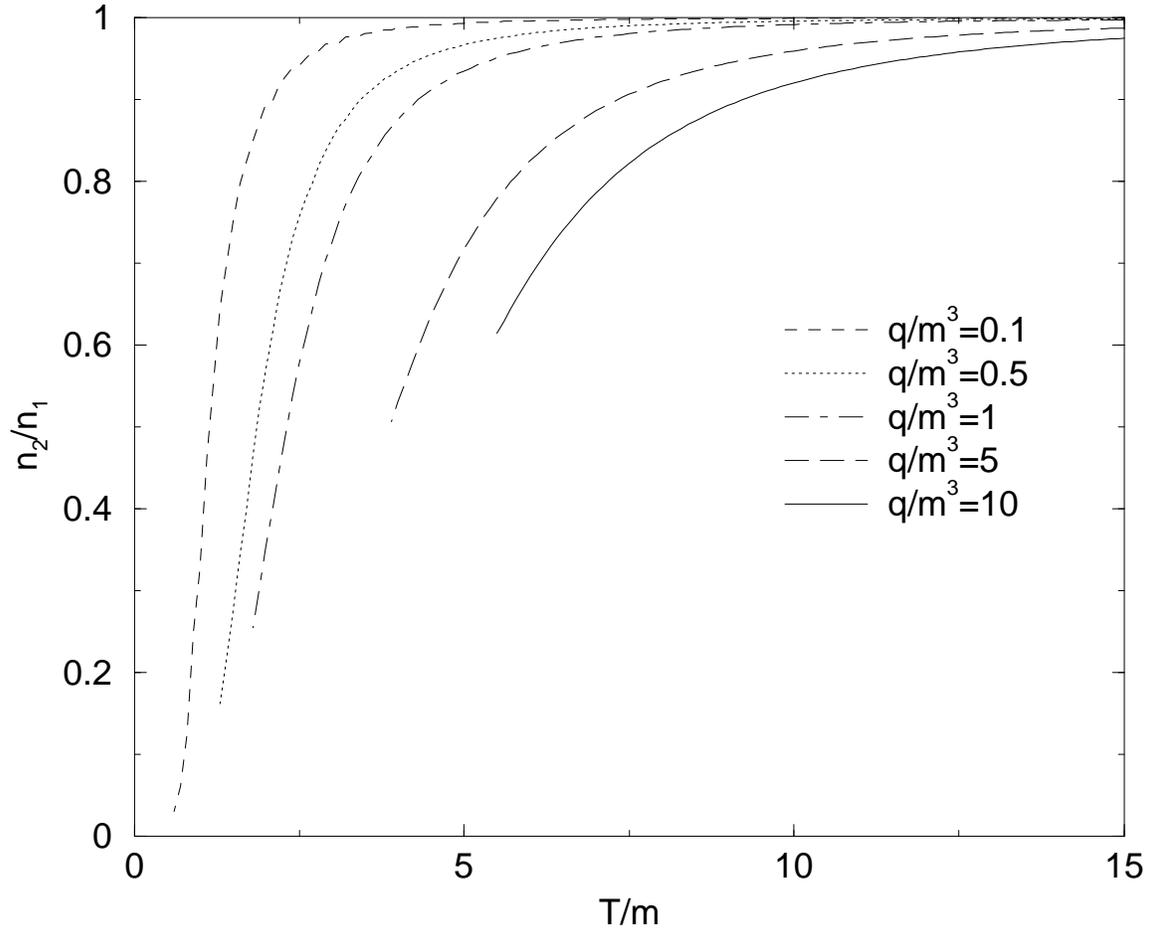,height=5.0in}}
\caption{Density ratio $n_2/n_1$ between particles and anti-particles 
{\it vs} scaled temperature $T/m$. Curves for different values 
of the scaled charge density $q/m^3$ of the Bose gas above 
the critical teperature $T_c$, where each curve ends.}
\end{figure}

\newpage

\begin{figure}
\centerline{\psfig{file=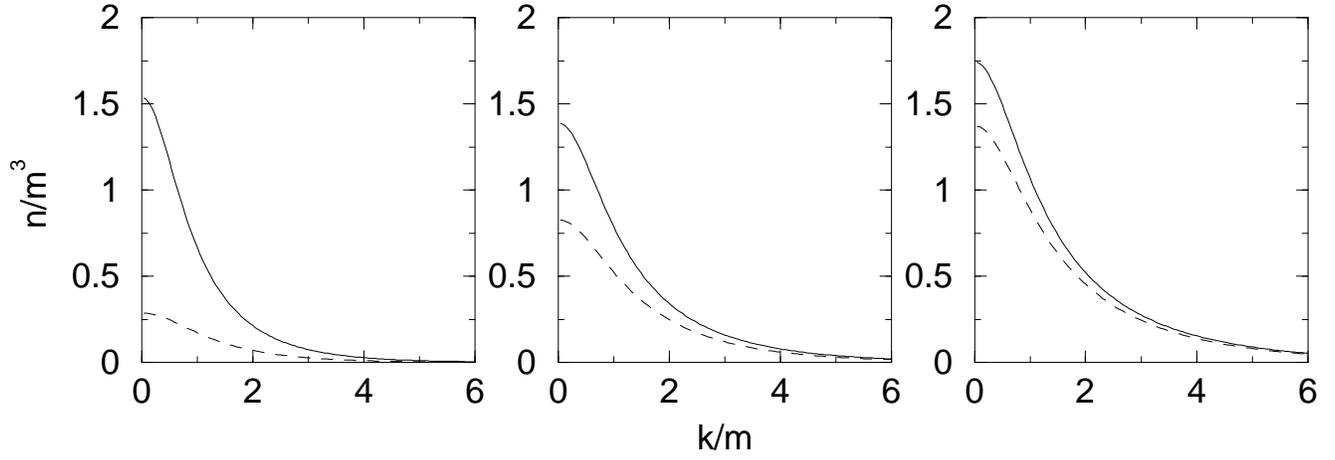,height=2.5in}}
\caption{Density profile in momentum space for particles 
(full line) and anti-particles (dashed line). 
Scaled charge density of the Bose gas: 
$q/m^3=0.1$. From left to right: $T/m=1$, $T/m=1.5$, $T/m=2$.}
\end{figure}

\newpage

\begin{figure}
\centerline{\psfig{file=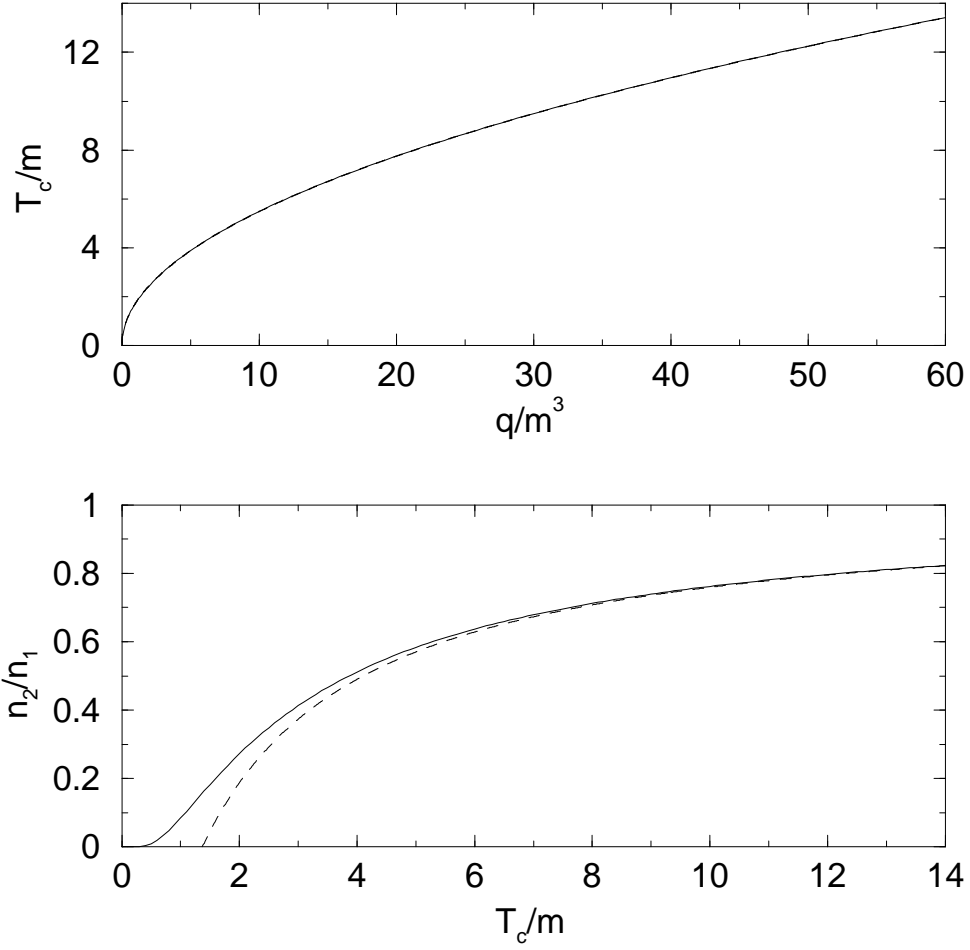,height=5in}}
\caption{
Universal curves at the BEC transition point. 
Scaled critical temperature $T_c/m$ {\it vs} scaled 
charge density $q/m^3$ (above). Density ratio $n_2/n_1$ 
{\it vs} scaled critical temperature $T_c/m$ (below). 
Full lines are numerical results and dashed lines 
are analytical results in the ultra-relativistic limit. 
}
\end{figure}

\newpage

\begin{figure}
\centerline{\psfig{file=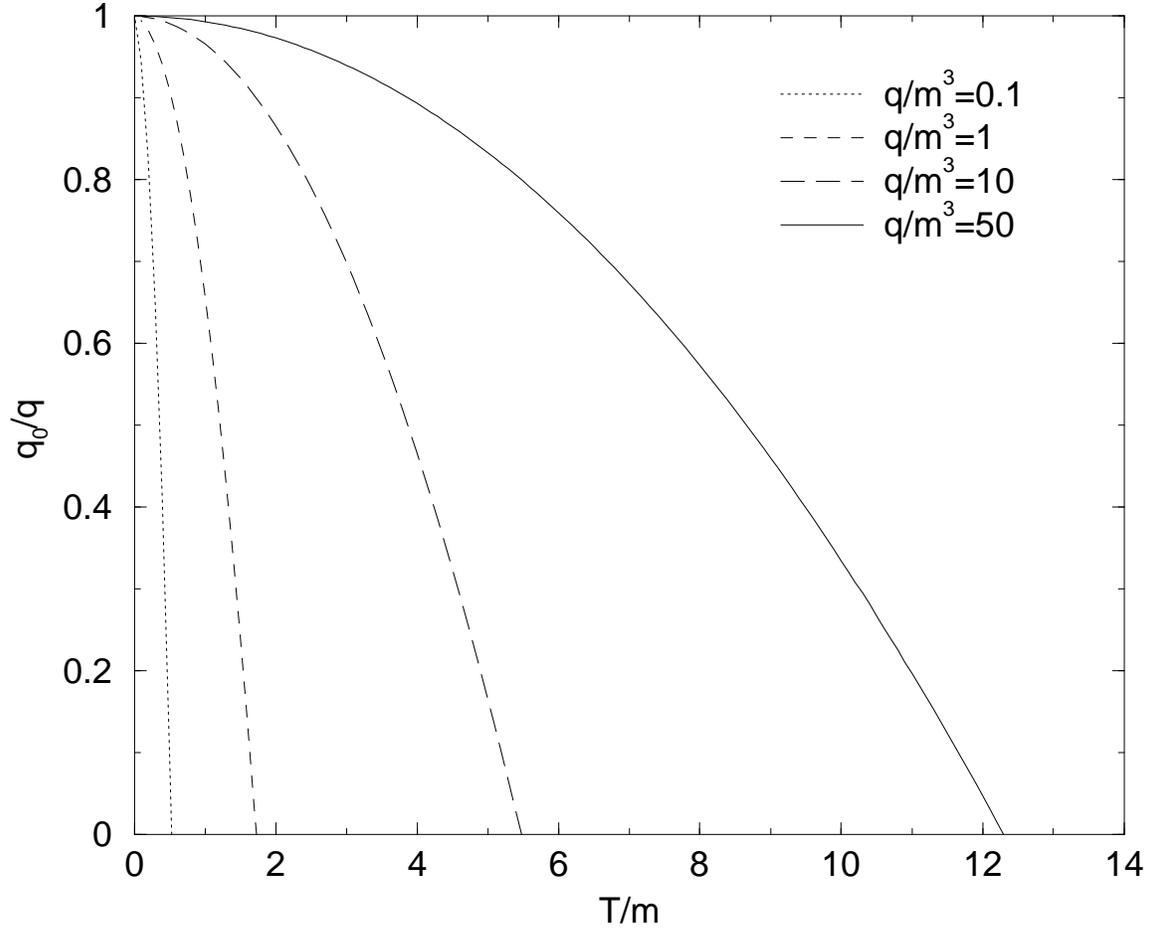,height=5in}}
\caption{
Condensed fraction $q_0/q$ {\it vs} scaled temperature $T/m$. 
Curves for different values of the scaled charge density $q/m^3$ of 
the Bose gas below the critical temperature.}
\end{figure}

\end{document}